# Orbit optimization and time delay interferometry for inclined ASTROD-GW formation with half-year precession-period

Wang Gang[a], and Ni Wei-Tou[b]

[a] *Shenzhen National Climate Observatory, No.1 Qixiang Rd., Zhuzilin, Futian District, Shenzhen, 518040, China*

[b] *Center for Gravitation and Cosmology (CGC), Department of Physics, National Tsing Hua University, Hsinchu, Taiwan, 30013, ROC*

e-mails: gwanggw@gmail.com, weitou@gmail.com

## Abstract

ASTROD-GW (ASTROD [Astrodynamical Space Test of Relativity using Optical Devices] optimized for Gravitational Wave detection) is a gravitational-wave mission with the aim of detecting gravitational waves from massive black holes, extreme mass ratio inspirals (EMRIs) and galactic compact binaries, together with testing relativistic gravity and probing dark energy and cosmology. Mission orbits of the 3 spacecrafts forming a nearly equilateral triangular array are chosen to be near the Sun-Earth Lagrange points L3, L4 and L5. The 3 spacecrafts range interferometrically with one another with arm length about 260 million kilometers. For 260 times longer arm length, the detection sensitivity of ASTROD-GW is 260 fold better than that of eLISA/NGO in the lower frequency region by assuming the same acceleration noise. Therefore, ASTROD-GW will be a better cosmological probe. In previous papers, we have worked out the time delay interferometry (TDI) for the ecliptic formation. To resolve the reflection ambiguity about the ecliptic plane in source position determination, we have changed the basic formation into slightly inclined formation with half-year precession-period. In this paper, we optimize a set of 10-year inclined ASTROD-GW mission orbits numerically using ephemeris framework starting at June 21, 2035, including cases of inclination angle is 0° (no inclination), 0.5°, 1.0°, 1.5°, 2.0°, 2.5° and 3.0°. We simulate the time delays of the first and second generation TDI configurations for the different inclinations, and compare/analyse the numerical results to attain the requisite sensitivity of ASTROD-GW by suppressing laser frequency noise below the secondary noises. To explicate our calculation process for different inclination cases, we take the 1.0° as example to show the orbit optimization and TDI simulation.

## 1. Introduction

Gravitational wave (GW) detection will provide a new method to probe astrophysics, cosmology and fundamental physics. The experimental GW communities in the world are actively looking forward to the rise of experimental GW astronomy and cosmology.[1,2] Cosmic Microwave Background (CMB) polarization observation teams are actively looking for B-mode polarization imprints from primordial/inflationary GWs.[3-7] The real-time GW detectors at present are mainly ground-based interferometers and Pulsar Timing Arrays (PTAs). The second-generation ground-based interferometers seeking to detect GWs in the high frequency band under active construction are Advanced LIGO,[8] Advanced Virgo[9] and KAGRA/LCGT.[10] LIGO-India is under active consideration.[11,12] PTAs seek for detection of GWs from supermassive black hole (SMBH) merger events and stochastic background in the very low frequency band (300 pHz – 100 nHz).[13-16] Prospects for real-time detection of GWs before or around 2020 are promising.

To detect GWs from various different astrophysical and cosmological sources in different spectral ranges and to enhance the signal to noise ratios, we need to explore the GW spectrum between the high frequency band and the very low frequency band, i.e., the middle frequency band (0.1 Hz – 10 Hz) and the low frequency band (100 nHz – 0.1 Hz).[1,2,17] Space detectors are most sensitive to these bands. Mission concepts under implementation and study are eLISA/NGO,[18] ASTROD-GW,[1,17,19-21] Super-ASTROD,[22] BBO[23] and DECIGO.[24-26] Except for the Fabry-Perot implementation of DECIGO whose scheme is in principle like the ground GW interferometric detectors, all other missions have unequal arm lengths and the laser frequency noise needs very serious consideration. One way to suppress it is to use time delay interferometry (TDI) by combining paths to make the two interfering beam to have closely equal optical paths. Time delay interferometry is considered for ASTROD in 1996[27,28] and has been worked out for LISA much more thoroughly since 1999.[29,30]

In our recent papers, we have worked out various TDI configurations numerically for ASTROD-GW, [31-34] LISA[35] and eLISA/NGO[36] and compared them with one another.[21] These second generation TDIs all satisfy their respective requirements. In the previous ASTROD-GW orbit configuration, we have used the ecliptic plane orbit formation in the original proposal. The angular position in the sky has a binary reflection ambiguity above and below the ecliptic plane. The resolution is poor when the source is near the ecliptic poles. Subsequently we have redesigned the basic orbits of ASTROD-GW configuration to have small inclinations with respect to the ecliptic plane to resolve these issues while keeping the variation of the arm lengths in the acceptable range.[17,21] In the present paper, we work out a set of 10-year optimized ASTROD-GW mission orbit configurations using numerical ephemeris, and calculate the residual optical path differences in the first and second generation TDIs for these inclined configurations. In our optimized mission orbits of ASTROD-GW for 10 years with 0° (no inclination), 0.5°, 1.0°, 1.5°, 2.0°, 2.5°, 3.0° inclined angles, changes of arm length are less than 0.0006, 0.0006, 0.0007, 0.0010, 0.0013, 0.0018, 0.0024 AU respectively; the Doppler velocities between each two S/Cs of the three arms are less than 3, 4, 10, 20, 34, 51, 73 m/s respectively; the rms (root-mean-square) Doppler velocities are about 1.2, 1.7, 5.6, 12.5, 22, 35, 50 m/s respectively. All the second generation TDIs considered for one-detector case with no inclination and with 0.5° inclination satisfy the original ASTROD-GW requirement on TDI path difference of less than 1.5 μs. For the case of 1.0° inclination, there are 10 second generation TDIs satisfy the requirement of mission in the 14 configurations calculated. For the cases of 1.5° to 3°, the requirement needs to be relaxed by 6 to 25 times; that is, the laser frequency stabilization noise needs to be suppressed by this additional factor. Among the first generation TDIs considered, the requirement for unequal arm Michelson, Relay, Beacon and Monitor needs to be relaxed by 1-3 orders. With the present pace of development, the laser frequency stabilization requirement for space equipment is expected to be able to compensate for this TDI requirement relaxation 20 years later.

In section 2, we review the basic inclined ASTROD-GW formation with half-year precession-period. In section 3, we use the CGC 2.7.1 ephemeris framework to optimize the mission orbit design numerically for ASTROD-GW formation to have nearly equal arm lengths and to have minimal line-of-site Doppler velocity between different pairs of spacecrafts for 10 years, including the cases of inclination angle 0.5°, 1°, 1.5°, 2°, 2.5° and 3° with respect to the ecliptic plane. In November 2013, ESA announced the selection of the Science Themes for the L2 and L3 launch opportunities -- the "Hot and Energetic Universe" for L2 and "The Gravitational Universe" for L3.[37] ESA L3 mission is likely to have a launch opportunity in 2034.[37] Since eLISA/NGO GW mission concept is the major candidate at this time and it takes one year to transfer to the science orbit, a starting time for science phase is likely in 2035. We take the starting time at June 21, 2035 to conform to the general trend of expectation of GW missions for the orbit design and the TDI simulation of science phase of ASTROD-GW with inclined orbit configurations. Since the design and simulation is for 10 years, if the starting time is before 2040, it could be used as a design and simulation for a five-year mission or more. To facilitate comparison, we also work out orbit configuration with no inclination at this starting time. In section 4, we calculate the TDIs for these ASTROD-GW orbit configurations. Section 4.1 summarizes the basics of time-delay interferometry. In section 4.2, we work out the first generation TDIs for ASTROD-GW numerically. In section 4.3, we work out the second generation TDIs for ASTROD-GW with one interferometer and two arms. In section 5, we conclude this paper with discussions.

## 2. Basic orbit design of the ASTROD-GW formation with inclination

In this section, we review and summarize the basic inclined mission formation for ASTROD-GW.

The mission configuration of ASTROD-GW is to have its 3 spacecrafts to form a nearly equilateral triangular array with each S/C near one of the Sun-Earth Lagrange points L3, L4 and L5 respectively focusing on GW detection at low frequency. In the original proposal, the ASTROD-GW orbits are chosen in the ecliptic plane, i.e. the inclination $\lambda = 0$ (no inclination). For GW detection, the source angular position determination in the sky has reflection ambiguity above and below the ecliptic plane. The original ASTROD-GW proposal has poor angular resolution near the ecliptic poles, although the resolution is good for the most of sky direction. Now we have redesigned the basic orbits of ASTROD-GW to have small inclinations to resolve these issues while keeping the variation of the arm lengths in the acceptable

range.[17,21]

The basic idea of the redesign is as follows. If the orbits of the ASTROD-GW spacecraft are inclined with a small angle λ, the interferometry plane with appropriate design is also inclined with similar angle. When the ASTROD-GW formation evolves, the interferometry plane can be designed to precess in the ecliptic solar-system barycentric frame with half-year period. For continuous or quasi-continuous GW sources both near the polar region and off the polar region, the angular positions will be resolved without reflection ambiguity.

The basic configuration uses inclined circular orbit in the heliocentric ecliptic coordinate system. The orbit equation for an inclined circular orbit is[17,21]

$$\begin{pmatrix} x' \\ y' \\ z' \end{pmatrix} = \begin{pmatrix} a\left[1-\sin^2 \Phi_0 \left(1-\cos\lambda\right)\right]\cos\varphi + a\sin\Phi_0\cos\Phi_0\left(1-\cos\lambda\right)\sin\varphi \\ a\cos\Phi_0\sin\Phi_0\left(1-\cos\lambda\right)\cos\varphi + a\left[1-\cos^2\Phi_0\left(1-\cos\lambda\right)\right]\sin\varphi \\ -a\sin\Phi_0\sin\lambda\cos\varphi + a\ \cos\Phi_0\sin\lambda\sin\varphi \end{pmatrix} \quad (1)$$

where λ is the inclination angle, $\Phi_0$ is the right ascension of ascending node (RAAN), and $\phi = \omega t + \phi_0$ with $\phi - \Phi_0$ the true anomaly and $a$ the semi-major axis corresponding to the mean motion $\omega$ of 1 rev/sidereal year.[17,21]

For the three orbits with inclination λ (in radian) in the convention used in this paper (the numbering of S/C II and S/C III is switched compared with that of Ref.'s [17, 27], i.e. S/C II ⟷ S/C III; this applies to all the formulas.), we have:

$$\begin{aligned} &\text{S/C I}: \Phi_0\left(\text{I}\right) = 270°,\ \ \varphi_0\left(\text{I}\right) = 0°; \\ &\text{S/C II}: \Phi_0\left(\text{II}\right) = 30°,\ \ \varphi_0\left(\text{II}\right) = 240°; \\ &\text{S/C III}: \Phi_0\left(\text{III}\right) = 150°,\ \ \varphi_0\left(\text{III}\right) = 120°. \end{aligned} \quad (2)$$

Defining

$$\xi \equiv 1 - \cos\lambda = 0.5\lambda^2 + \text{O}\left(\lambda^4\right), \quad (3)$$

we have explicitly

(i)    for the orbit of S/C I

$$\begin{pmatrix} x^{\mathrm{I}} \\ y^{\mathrm{I}} \\ z^{\mathrm{I}} \end{pmatrix} = \begin{pmatrix} a\cos\omega t - a\xi\cos\omega t \\ a\sin\omega t \\ a\cos\omega t\sin\lambda \end{pmatrix}, \tag{4}$$

(ii)   for the orbit of S/C II

$$\begin{pmatrix} x^{\mathrm{II}} \\ y^{\mathrm{II}} \\ z^{\mathrm{II}} \end{pmatrix} = \begin{pmatrix} a[(-1/2)\cos\omega t - (3^{1/2}/2)\sin\omega t] + (a/2)\xi[(3^{1/2}/2)\sin\omega t - (1/2)\cos\omega t] \\ a[(-1/2)\sin\omega t + (3^{1/2}/2)\cos\omega t] + (3^{1/2}/2)a\xi[(3^{1/2}/2)\sin\omega t - (1/2)\cos\omega t] \\ a\sin\lambda[(3^{1/2}/2)\sin\omega t - (1/2)\cos\omega t] \end{pmatrix}. \tag{5}$$

(iii)   for the orbit of S/C III

$$\begin{pmatrix} x^{\mathrm{III}} \\ y^{\mathrm{III}} \\ z^{\mathrm{III}} \end{pmatrix} = \begin{pmatrix} a[(-1/2)\cos\omega t + (3^{1/2}/2)\sin\omega t] + (a/2)\xi[(3^{1/2}/2)\sin\omega t - (1/2)\cos\omega t] \\ a[(-1/2)\sin\omega t - (3^{1/2}/2)\cos\omega t] - (3^{1/2}/2)a\xi[(-3^{1/2}/2)\sin\omega t - (1/2)\cos\omega t] \\ a\sin\lambda[(-3^{1/2}/2)\sin\omega t - (1/2)\cos\omega t] \end{pmatrix}. \tag{6}$$

Fig. 1 shows the orbits in 3 dimension for inclination $\lambda = 1°$ case with the scale of $z$-axis blowing up by tenfold.

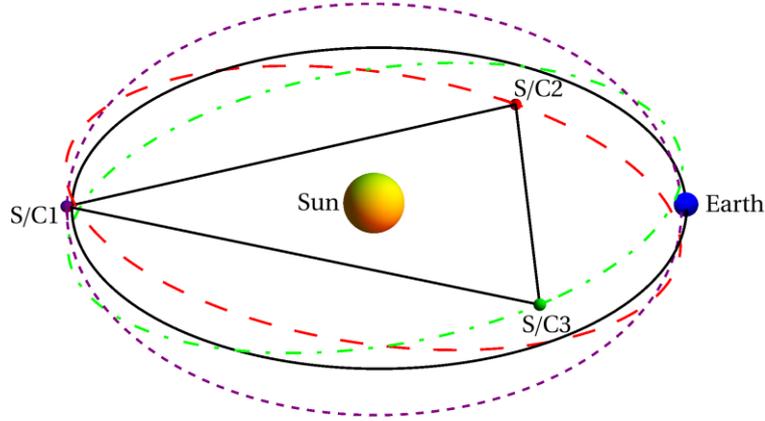

**Fig.1.** Schematic of ASTROD-GW mission orbit design with inclination.

With equations (4)-(6), the arm lengths are calculated to be

$$| \boldsymbol{V}_{\mathrm{II\text{-}I}} | = 3^{1/2}a\left[ \left(1 - \xi/2\right)^2 + \sin^2\lambda\sin^2\left(\omega t - 60°\right) \right]^{1/2},$$

$$| \boldsymbol{V}_{\mathrm{III\text{-}II}} | = 3^{1/2}a\left[ \left(1 - \xi/2\right)^2 + \sin^2\lambda\sin^2\left(\omega t\right) \right]^{1/2}, \tag{7}$$

$$| \boldsymbol{V}_{\mathrm{I\text{-}III}} | = 3^{1/2}a\left[ \left(1 - \xi/2\right)^2 + \sin^2\lambda\sin^2\left(\omega t + 60°\right) \right]^{1/2}.$$

The fractional arm length variation is within $\pm (1/2)\sin^2\lambda$, e.g. $\pm 0.76 \times 10^{-4}$ for $\lambda = 1°$.

The Doppler velocity (line-of-sight velocity) between two spacecrafts, e.g. S/C II

and S/C III is

$$\frac{\mathrm{d}\,|\,\boldsymbol{V}_{\mathrm{III\text{-}II}}\,|}{\mathrm{d}t} = (3^{1/2}/2)a\omega\sin^2\lambda\sin(2\omega t)\left[(1-\xi/2)^2 + \sin^2\lambda\sin^2(\omega t)\right]^{-1/2}. \quad (8)$$

From this equation the line-of-sight Doppler velocity is less than $(3^{1/2}/2)\,a\omega\,\sin^2\lambda\,(1 + O(\lambda^2))$.

The fractional arm length variations, arm length variations and Doppler velocity variations for the basic configurations of ASTROD-GW for 10 years respectively with the inclination angle 0° (no inclination), 0.5°, 1°, 1.5°, 2°, 2.5° and 3° to the ecliptic plane are listed in columns 2-4 in Table 1. In section 3 we do orbit optimization to these formations with starting time at June 21, 2035 (JD2461944.0) including planetary perturbations. Their values of Doppler velocity variations and arm length variations in 10 years are listed in columns 5-6 of Table 1.

**Table 1.** Inclined ASTROD-GW formations, arm length variation and Doppler velocity variations

| Nominal Orbit Inclination $\lambda$ | Fractional arm length variation (2 body gravitational field) | Arm length variation (2 body gravitational field) | Doppler velocity variation (2 body gravitational field) | Simulated Doppler velocity variation with planetary perturbations (10 years) | Simulated arm length variation with planetary perturbations (10 years) |
|---|---|---|---|---|---|
| 0° | 0 | 0 | 0 | 3 m/s | $6 \times 10^{-4}$ AU |
| 0.5° | $0.19 \times 10^{-4}$ | $0.33 \times 10^{-4}$ AU | 2.0 m/s | 4 m/s | $6 \times 10^{-4}$ AU |
| 1° | $0.76 \times 10^{-4}$ | $1.32 \times 10^{-4}$ AU | 7.9 m/s | 10 m/s | $7 \times 10^{-4}$ AU |
| 1.5° | $1.71 \times 10^{-4}$ | $2.97 \times 10^{-4}$ AU | 18 m/s | 20 m/s | $10 \times 10^{-4}$ AU |
| 2° | $3.04 \times 10^{-4}$ | $5.27 \times 10^{-4}$ AU | 32 m/s | 34 m/s | $13 \times 10^{-4}$ AU |
| 2.5° | $4.75 \times 10^{-4}$ | $8.23 \times 10^{-4}$ AU | 49 m/s | 51 m/s | $18 \times 10^{-4}$ AU |
| 3° | $6.84 \times 10^{-4}$ | $11.8 \times 10^{-4}$ AU | 71 m/s | 73 m/s | $24 \times 10^{-4}$ AU |

The cross-product vector $\boldsymbol{N}(t) \equiv \boldsymbol{V}_{\text{III-II}} \times \boldsymbol{V}_{\text{I-III}}$ is normal to the orbit configuration plane and has the following components:

$$\boldsymbol{N} = (3^{3/2} / 2) a^2 \left(1 - \xi / 2\right) \begin{pmatrix} -\sin \lambda \cos 2\omega t \\ -\sin \lambda \sin 2\omega t \\ 1 - \xi / 2 \end{pmatrix}. \tag{9}$$

The normalized unit normal vector $\boldsymbol{n}$ is then:

$$\boldsymbol{n} = \left[ \sin^2 \lambda + \left(1 - \xi / 2\right)^2 \right]^{-1/2} \begin{pmatrix} -\sin \lambda \cos 2\omega t \\ -\sin \lambda \sin 2\omega t \\ 1 - \xi / 2 \end{pmatrix}. \tag{10}$$

Thus, the ASTROD-GW formation precesses with angular velocity $2\omega$, i.e. with precession period half sidereal year.

The geometric center $\boldsymbol{V}_{\text{c}}$ of the ASTROD-GW spacecraft configuration is

$$\boldsymbol{V}_{\text{c}} = \begin{pmatrix} -(1 / 2) a \xi \cos \omega t \\ (1 / 2) a \xi \sin \omega t \\ 0 \end{pmatrix}. \tag{11}$$

There are 3 interferometers with 2 arms in the ASTROD-GW configuration. The geometric centers of these 3 interferometers are at a distance of about 0.25 AU from the Sun. For 6 inclinations between 0.5° to 3°, we optimize their orbit configurations and simulate various TDI numerically using planetary ephemeris to take into account the planetary perturbations. When eLISA formation orbits around the Sun, it is equivalent to multiple detector arrays distributed in 1 AU orbit. The extension of ASTROD-GW is already of 1 AU. When ASTROD-GW formation orbits around the Sun, it is also equivalent to multiple detector arrays distributed in 1 AU orbit. For its angular resolution compared with eLISA, see Ref's [17, 21].

## 3. Mission orbit optimization

The goal of ASTROD-GW mission orbit optimization is to equalize the three arm lengths of ASTROD-GW formation and to reduce the relative line-of-sight velocities between three pairs of spacecrafts as much as possible in realistic situation with planetary perturbations. In the solar system, the periods and eccentricities of the ASTROD-GW spacecraft orbits are perturbed by the other planets with largest caused by Jupiter and Venus. Our method of optimization is to modify the initial velocities and initial heliocentric distances so that the perturbed orbital periods for ten-year average remain close to a sidereal year and the average eccentricities remain near zero. In our first optimization, the start time of the science part of the mission is chosen to be June 21, 2025 (JD2460848.0) and the optimization is for a period of 3700 days.[38,39] Since the preparation of the mission may take longer time, in our second optimization, we start at noon, June 21, 2028 (JD2461944.0) and optimize for a period of 20 years for doing numerical TDIs.[31-34] In this paper, we optimize the orbit configurations of ASTROD-GW for 10 years for the cases with the inclination angle 0.5°,1°, 1.5°, 2°, 2.5° and 3° to the ecliptic plane starting at noon, June 21, 2035 (JD2464500.0) for reason explained in the Introduction. For reference, we also do the optimization for no inclination case.

We calculate and optimize the realistic orbit configurations by using the CGC 2.7.1 ephemeris. In our previous paper,[34] we used the CGC 2.7 ephemeris. The differences between CGC 2.7.1 and CGC 2.7 is detailed in subsection 3.1. In subsection 3.2, we obtain the initial choice of S/C initial conditions as a starting point for optimization. In subsection 3.3, we discuss method of optimization. In subsection 3.4, we present the results of optimization.

## 3.1 CGC 2.7.1 Ephemeris

Modern ephemerides are built upon the post-Newtonian dynamics.[40-42] The CGC ephemeris use the post-Newtonian equations given by Brumberg.[43] In the CGC 2.7.1 ephemeris framework, we pick up 340 asteroids besides the Ceres, Pallas and Vesta from the Lowell database.[44] The masses of 340 asteroids are given by Lowell data[44] instead of estimating the masses based on the classification in CGC 2.7.[34-36] The orbit elements of these asteroids are also updated from the Lowell database.

For a 10-year duration starting at June 21, 2035, the differences between the Earth's heliocentric distances calculated by CGC 2.7.1 and DE430 are within 150 m, and that the differences in longitudes and latitudes are within 1.4 mas and 0.45 mas respectively. These differences will not affect the results of our TDI calculations.

### 3.2 Initial choice of spacecraft initial conditions

The R.A. of the Earth at JD2464500 (2035-June-21st 12:00:00) is $17^h57^m45.09^s$, i.e. $269.438°$ from DE 430 ephemeris. The initial positions of the 3 S/Cs are obtained by choosing the $\omega t$ as $89.44°$ for $\phi = \omega t + \phi_0$ in Equation (1). The initial velocities are derived from Equation (1) by calculating the derivatives with respect to $t$. The S/C1 near the Lagrange point L3 is partly obscured by Sun from the line of sight of Earth (left diagram of Fig. 2). It would obstruct the communication with the Earth stations. To avoid the obscuration, we rotate the initial angle $\Phi_0$ and $\phi_0$ forward of by $2.8°$, $2.4°$, $2.0°$, $1.8°$, $1.4°$, $1.3°$ and $1.2°$ for inclination angle $0.0°$, $0.5°$, $1.0°$, $1.5°$, $2.0°$, $2.5°$ and $3.0°$ respectively. For the case with inclination angle $1.0°$, the S/C1 orbit is shown on the right diagram of Fig. 2. The initial choice of initial states for the 3 S/Cs in this case is listed in column 3 of Table 2.

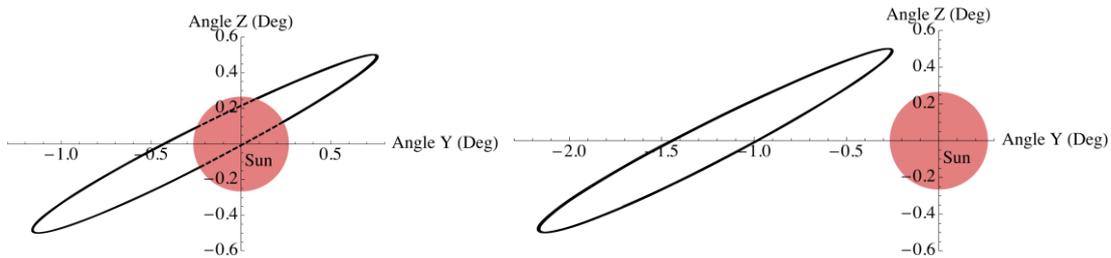

**Fig. 2.** S/C1 view form Earth before rotating the initial conditions by an angle (left diagram) and after rotating by an angle $2.0°$ (right diagram) for the case of inclination angle $1.0°$.

**Table 2.** Initial states of S/Cs for the configuration with the inclination angle 1° at epoch JD2464500.0 for initial choice, after period optimization, and after all optimizations in J2000 equatorial solar-system-barycentric coordinate system.

| $\lambda = 1.0°$ | | Initial choice of S/Cs initial states | Initial states of S/Cs after period optimization | Initial states of S/Cs after final optimization |
|---|---|---|---|---|
| S/C1 | X | $-2.8842263289715\times10^{-2}$ | $-2.8842263289715\times10^{-2}$ | $-2.8842514605546\times10^{-2}$ |
| Position | Y | $9.1157742309044\times10^{-1}$ | $9.1157742309044\times10^{-1}$ | $9.1158659433458\times10^{-1}$ |
| (AU) | Z | $3.9552690922456\times10^{-1}$ | $3.9552690922456\times10^{-1}$ | $3.9553088730467\times10^{-1}$ |
| S/C1 | Vx | $-1.7188548244458\times10^{-2}$ | $-1.7188535691176\times10^{-2}$ | $-1.7188363750567\times10^{-2}$ |
| Velocity | Vy | $-2.8220395391983\times10^{-4}$ | $-2.8220375159556\times10^{-4}$ | $-2.8220098038726\times10^{-4}$ |
| (AU/day) | Vz | $-4.4970276654173\times10^{-4}$ | $-4.4970243993363\times10^{-4}$ | $-4.4969796642665\times10^{-4}$ |
| S/C2 | X | $8.7453598387569\times10^{-1}$ | $8.7453598387569\times10^{-1}$ | $8.7453598387569\times10^{-1}$ |
| Position | Y | $-4.3802677355114\times10^{-1}$ | $-4.3802677355114\times10^{-1}$ | $-4.3802677355114\times10^{-1}$ |
| (AU) | Z | $-2.0634980179207\times10^{-1}$ | $-2.0634980179207\times10^{-1}$ | $-2.0634980179207\times10^{-1}$ |
| S/C2 | Vx | $8.2301784322477\times10^{-3}$ | $8.2301033726700\times10^{-3}$ | $8.2301033726700\times10^{-3}$ |
| Velocity | Vy | $1.3797379424198\times10^{-2}$ | $1.3797253460590\times10^{-2}$ | $1.3797253460590\times10^{-2}$ |
| (AU/day) | Vz | $6.1425805519808\times10^{-3}$ | $6.1425244722884\times10^{-3}$ | $6.1425244722884\times10^{-3}$ |
| S/C3 | X | $-8.5683596527799\times10^{-1}$ | $-8.5683596527799\times10^{-1}$ | $-8.5679330969623\times10^{-1}$ |
| Position | Y | $-4.8998222347472\times10^{-1}$ | $-4.8998222347472\times10^{-1}$ | $-4.8995800210059\times10^{-1}$ |
| (AU) | Z | $-1.9592963105165\times10^{-1}$ | $-1.9592963105165\times10^{-1}$ | $-1.9591994878015\times10^{-1}$ |
| S/C3 | Vx | $8.9788714330506\times10^{-3}$ | $8.9787977300014\times10^{-3}$ | $8.9792464008067\times10^{-3}$ |
| Velocity | Vy | $-1.3530263187520\times10^{-2}$ | $-1.3530152097744\times10^{-2}$ | $-1.3530828362023\times10^{-2}$ |
| (AU/day) | Vz | $-5.6998631854817\times10^{-3}$ | $-5.6998163886731\times10^{-3}$ | $-5.7001012664635\times10^{-3}$ |

### 3.3 Method of optimization

Our optimization method is to modify the initial velocities and initial heliocentric distances to reach the aim of (i) equalizing the three arm lengths of the ASTROD-GW formation as much as possible and (ii) reducing the relative Doppler velocities between three pairs of spacecrafts as much as possible.

During the actual optimization procedure, we use the following equation to modify the average period of the orbit:

$$\boldsymbol{V}_{new} = \boldsymbol{V}_{prev} + \Delta\boldsymbol{V} \approx (1 - \frac{1}{3}\frac{\Delta T}{T})\boldsymbol{V}_{prev} \tag{12}$$

For the case of inclination angle of 1°, we calculate the 3 S/Cs orbits with the initial choice of initial conditions listed in column 3 of Table 2 using the CGC 2.7.1 ephemeris. The average periods of the 3 S/Cs in 10 years are 365.256 days (S/C1), 365.267 days (S/C2) and 365.266 days (S/C3) respectively. We use equation (12) to change the initial velocities so that the average period of S/C1, S/C2 and S/C3 is adjusted to 365.255 days, 365.257 days and 365.257 days respectively. The initial conditions after this step are listed in column 4 of Table 2. In the next step, we use the following equations to trim the S/C eccentricities to be nearly circular:

$$
\begin{aligned}
\boldsymbol{R}_{\text{new}} &= \boldsymbol{R}_{\text{prev}} + \Delta \boldsymbol{R} \approx (1 + \frac{\Delta R}{R})\boldsymbol{R}_{\text{prev}} \\
\boldsymbol{V}_{\text{new}} &= \boldsymbol{V}_{\text{prev}} + \Delta \boldsymbol{V} \approx (1 - \frac{\Delta R}{R})\boldsymbol{V}_{\text{prev}}
\end{aligned}
. \tag{13}
$$

Here $R$ is the initial heliocentric distance of spacecraft. The fractional adjustment $\pm(\Delta R/R)$ in $\boldsymbol{R}_{prev}$ and $\boldsymbol{V}_{prev}$ would adjust eccentricity without adjust the period of the orbit. The initial conditions after all optimization are listed in column 5 of Table 2.

As for the inclination angle is 0.0°, 0.5°, 1.5°, 2°, 2.5° and 3°, the optimization processes are similar to the inclination is 1.0°.

In Fig. 3, the variation of (a) arm lengths, (b) difference of arm lengths, (c) angles between arms, (d) velocities in the measure direction, (e) inclination of the unit normal $\boldsymbol{n}$ of the ASTROD-GW formation and (f) azimuthal angle of $\boldsymbol{n}$ in 10 years are illustrated for the case with the orbit inclination angle 1°. For this and other inclination cases, their values of Doppler velocity variations and arm length variations in 10 years are listed in columns 5-6 of Table 1 for comparison.

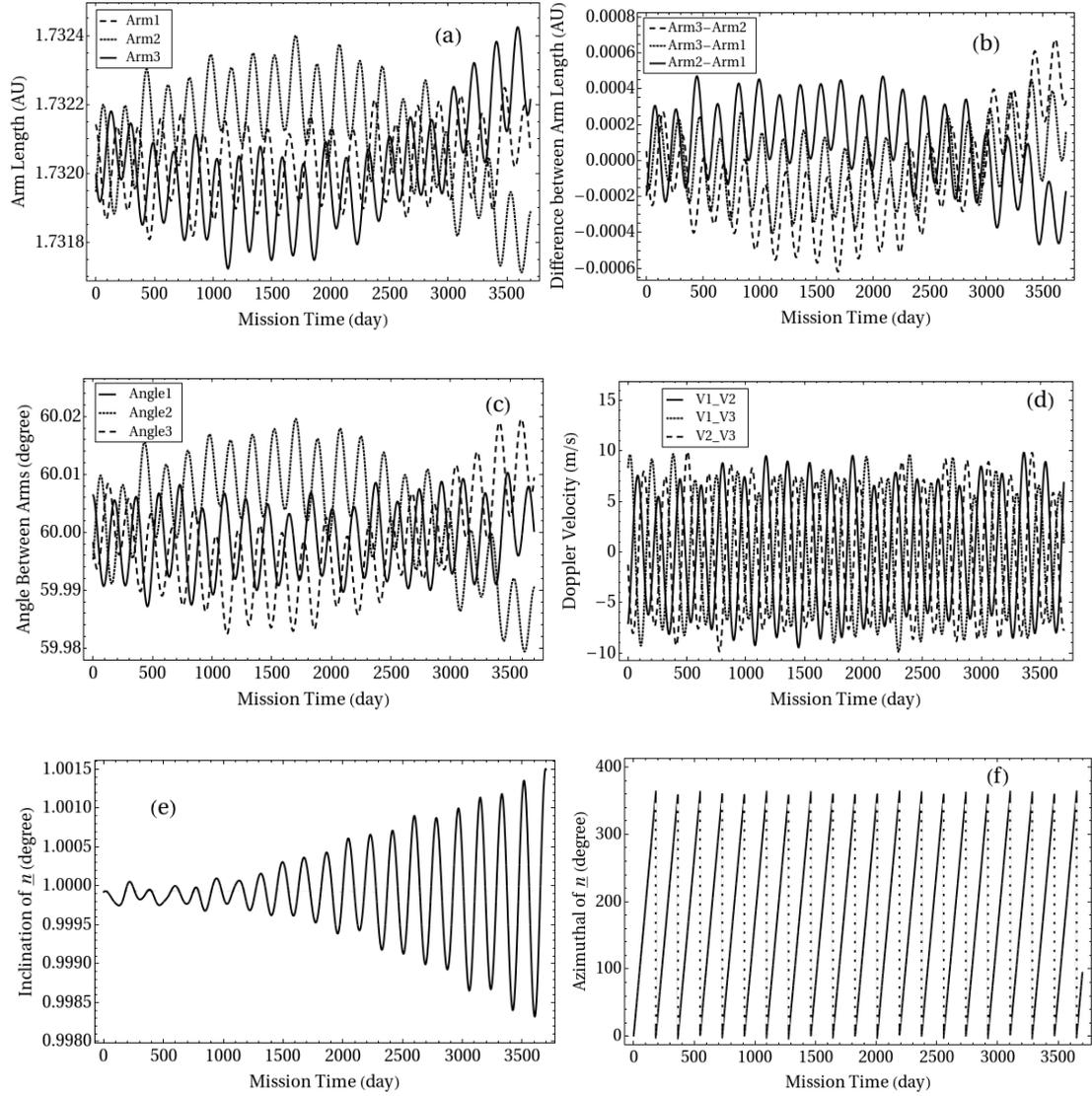

**Fig. 3.** The variation of (a) arm lengths, (b) difference of arm lengths, (c) angles between arms, (d) velocities in the measure direction, (e) inclination of the unit normal ***n*** of the ASTROD-GW formation and (f) azimuthal angle of ***n*** in 10 years for the case with the orbit inclination angle 1°.

### 3.4 Results of optimization

The initial positions and initial velocities of the 3 spacecrafts obtained after the optimization processes of Section 3.3 in J2000 equatorial solar-system barycentric coordinate system for the inclination angle 0°, 0.5°, 1°, 1.5°, 2°, 2.5° and 3° are listed in the third-to-fifth columns of Table 3.

**Table 3.** Initial states of S/Cs at epoch JD2464500.0 for the 3 S/Cs the after optimizations in J2000 equatorial solar-system-barycentric coordinate system for the inclination angle $\lambda = 0.0°, \ 0.5°, \ 1.0°, 1.5°, 2.0°, \ 2.5°$ and $3.0°$ respectively.

| $\lambda$ | | S/C1 | S/C2 | S/C3 |
|---|---|---|---|---|
| 0°; | X | $-4.2796878676817 \times 10^{-2}$ | $8.8117798070826 \times 10^{-1}$ | $-8.4947241483190 \times 10^{-1}$ |
| Initial position | Y | $9.1125233495679 \times 10^{-1}$ | $-4.3287312094977 \times 10^{-1}$ | $-4.9495904068582 \times 10^{-1}$ |
| (AU); | Z | $3.9520005050495 \times 10^{-1}$ | $-1.8754965452141 \times 10^{-1}$ | $-2.1446719503333 \times 10^{-1}$ |
| initial | Vx | $-1.7183095034183 \times 10^{-2}$ | $8.0177988365537 \times 10^{-3}$ | $9.1828142475163 \times 10^{-3}$ |
| velocity | Vy | $-6.2194110283150 \times 10^{-4}$ | $1.3961176877644 \times 10^{-2}$ | $-1.3355134639207 \times 10^{-2}$ |
| (AU/day) | Vz | $-2.6979249323306 \times 10^{-4}$ | $6.0527629379897 \times 10^{-3}$ | $-5.7903075201295 \times 10^{-3}$ |
| 0.5°; | X | $-3.5819911669671 \times 10^{-2}$ | $8.7790469169867 \times 10^{-1}$ | $-8.5317761553607 \times 10^{-1}$ |
| Initial position | Y | $9.1143723606865 \times 10^{-1}$ | $-4.3549057022393 \times 10^{-1}$ | $-4.9250178488352 \times 10^{-1}$ |
| (AU); | Z | $3.9537317729935 \times 10^{-1}$ | $-1.9696765577002 \times 10^{-1}$ | $-2.0521201708840 \times 10^{-1}$ |
| initial | Vx | $-1.7186887468334 \times 10^{-2}$ | $8.1238999308544 \times 10^{-3}$ | $9.0809973986794 \times 10^{-3}$ |
| velocity | Vy | $-4.5210095520829 \times 10^{-4}$ | $1.3879665203167 \times 10^{-2}$ | $-1.3443438367041 \times 10^{-2}$ |
| (AU/day) | Vz | $-3.5976439874011 \times 10^{-4}$ | $6.0978417340874 \times 10^{-3}$ | $-5.7453995292536 \times 10^{-3}$ |
| 1.0°; | X | $-2.8842514605546 \times 10^{-2}$ | $8.7453598387569 \times 10^{-1}$ | $-8.5679330969623 \times 10^{-1}$ |
| Initial position | Y | $9.1158659433458 \times 10^{-1}$ | $-4.3802677355114 \times 10^{-1}$ | $-4.8995800210059 \times 10^{-1}$ |
| (AU); | Z | $3.9553088730467 \times 10^{-1}$ | $-2.0634980179207 \times 10^{-1}$ | $-1.9591994878015 \times 10^{-1}$ |
| initial | Vx | $-1.7188363750567 \times 10^{-2}$ | $8.2301033726700 \times 10^{-3}$ | $8.9792464008067 \times 10^{-3}$ |
| velocity | Vy | $-2.8220098038726 \times 10^{-4}$ | $1.3797253460590 \times 10^{-2}$ | $-1.3530828362023 \times 10^{-2}$ |
| (AU/day) | Vz | $-4.4969796642665 \times 10^{-4}$ | $6.1425244722884 \times 10^{-3}$ | $-5.7001012664635 \times 10^{-3}$ |
| 1.5°; | X | $-2.5354637468848 \times 10^{-2}$ | $8.7277485656422 \times 10^{-1}$ | $-8.5853944803538 \times 10^{-1}$ |
| Initial position | Y | $9.1162752109802 \times 10^{-1}$ | $-4.3768521777563 \times 10^{-1}$ | $-4.9008352249095 \times 10^{-1}$ |
| (AU); | Z | $3.9564157234190 \times 10^{-1}$ | $-2.1448302499827 \times 10^{-1}$ | $-1.8778616669275 \times 10^{-1}$ |
| initial | Vx | $-1.7186475849607 \times 10^{-2}$ | $8.2836295551138 \times 10^{-3}$ | $8.9286603191558 \times 10^{-3}$ |
| velocity | Vy | $-1.6734353206691 \times 10^{-4}$ | $1.3740218938950 \times 10^{-2}$ | $-1.3588357937831 \times 10^{-2}$ |
| (AU/day) | Vz | $-5.6346990499641 \times 10^{-4}$ | $6.1981969674239 \times 10^{-3}$ | $-5.6418723212812 \times 10^{-3}$ |
| 2.0°; | X | $-1.8377019454199 \times 10^{-2}$ | $8.6928410315028 \times 10^{-1}$ | $-8.6202745951776 \times 10^{-1}$ |
| Initial position | Y | $9.1170983137539 \times 10^{-1}$ | $-4.4008950965408 \times 10^{-1}$ | $-4.8739932270530 \times 10^{-1}$ |
| (AU); | Z | $3.9577017809207 \times 10^{-1}$ | $-2.2380515792750 \times 10^{-1}$ | $-1.7843625650587 \times 10^{-1}$ |
| initial | Vx | $-1.7184087416603 \times 10^{-2}$ | $8.3897932228791 \times 10^{-3}$ | $8.8269105632561 \times 10^{-3}$ |
| velocity | Vy | $2.5668890281350 \times 10^{-6}$ | $1.3655647257599 \times 10^{-2}$ | $-1.3673677978699 \times 10^{-2}$ |
| (AU/day) | Vz | $-6.5333704018832 \times 10^{-4}$ | $6.2419097897323 \times 10^{-3}$ | $-5.5957087504276 \times 10^{-3}$ |

| 2.5; | X | $-1.6635186808372\times10^{-2}$ | $8.6826941613519\times10^{-1}$ | $-8.6277837376267\times10^{-1}$ |
|---|---|---|---|---|
| Initial position | Y | $9.1169794863181\times10^{-1}$ | $-4.3822782009220\times10^{-1}$ | $-4.8877342743721\times10^{-1}$ |
| (AU); | Z | $3.9585791785736\times10^{-1}$ | $-2.3127457630524\times10^{-1}$ | $-1.7084833096099\times10^{-1}$ |
| initial | Vx | $-1.7178593572115\times10^{-2}$ | $8.4175795794339\times10^{-3}$ | $8.8025653512540\times10^{-3}$ |
| velocity | Vy | $8.9879005834072\times10^{-5}$ | $1.3610739355664\times10^{-2}$ | $-1.3715709566895\times10^{-2}$ |
| (AU/day) | Vz | $-7.7896390115972\times10^{-4}$ | $6.3028004648275\times10^{-3}$ | $-5.5308031306408\times10^{-3}$ |
| 3.0°; | X | $-1.4894058497576\times10^{-2}$ | $8.6722736973803\times10^{-1}$ | $-8.6349420788078\times10^{-1}$ |
| Initial position | Y | $9.1168328063245\times10^{-1}$ | $-4.3631776883318\times10^{-1}$ | $-4.9009275125682\times10^{-1}$ |
| (AU); | Z | $3.9594441468004\times10^{-1}$ | $-2.3872020555868\times10^{-1}$ | $-1.6323977249558\times10^{-1}$ |
| initial | Vx | $-1.7171754206138\times10^{-2}$ | $8.4455810575003\times10^{-3}$ | $8.7785311071194\times10^{-3}$ |
| velocity | Vy | $1.7717529782666\times10^{-4}$ | $1.3565154450644\times10^{-2}$ | $-1.3757170437319\times10^{-2}$ |
| (AU/day) | Vz | $-9.0453594279848\times10^{-4}$ | $6.3633637953983\times10^{-3}$ | $-5.4656812863068\times10^{-3}$ |

## 4. Time-delay interferometry (TDI)

For laser-interferometric antenna for space detection of GWs, the arm lengths vary according to orbit dynamics. In order to attain the requisite sensitivity, laser frequency noise must be suppressed below the secondary noises such as the optical path noise, acceleration noise etc. For suppressing laser frequency noise, it's necessary to use time delay interferometry in the analysis to match the optical path length of different beams. The better match of the optical path lengths are, the better cancellation of the laser frequency noise and the easier to achieve the requisite sensitivity. In case of exact match, the laser frequency noise is fully cancelled, as in the original Michelson interferometer.

### 4.1. Basics of time delay interferometry

The TDI was first used in the study of ASTROD mission concept.[27,28,45] In the deep-space interferometry, long distances are invariably involved. Due to long distances, laser light is attenuated to a great extent at the receiving spacecraft. To transfer the laser light back or to another spacecraft, amplification is needed. The procedure is to phase lock the local laser to the incoming weak laser light and to transmit the local laser light back to another spacecraft. We have demonstrated the phase locking of a local oscillator with 2-pW laser light in laboratory.[46,47] Dick et al.[48] have demonstrated phase locking to 40-fW incoming weak laser light. The power requirement feasibility for ASTROD-GW is met with these developments. In the 1990s, we used the following two TDI configurations during the study of ASTROD interferometry and obtained numerically the path length differences using the Newtonian dynamics:[27,28]

   i) unequal arm Michelson TDI configuration

    Path 1: S/C1→S/C2→S/C1→S/C3→S/C1,

    Path 2: S/C1→S/C3→S/C1→S/C2→S/C1;

   ii) Sagnac TDI configuration:

    Path 1: S/C1→S/C2→S/C3→S/C1,

    Path 2: S/C1→S/C3→S/C2→S/C1.

Here we do the same thing for ASTROD-GW with inclined orbit. For the numerical evaluation, we take a common receiving time epoch for both beams; the results would be very close to each other numerically if we take the same start time epoch and calculate the path differences. The results of this calculation for orbit configuration with 1° inclination are shown in Fig. 4. We refer to the path S/C1→S/C2→S/C1 as $a$ and the path S/C1 → S/C3 → S/C1 as $b$. Hence the difference $\Delta L$ between Path 1 and Path 2 for the unequal-arm Michelson can be denoted as $ab - ba \equiv [a,b]$.

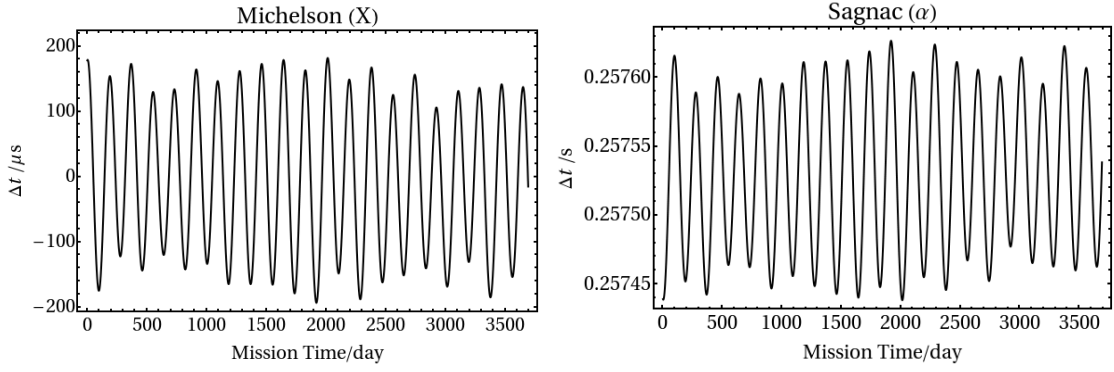

**Fig. 4.** Path length differences between two optical paths of the Unequal-arm Michelson TDI configuration (X) and Sagnac TDI configuration (α) for ASTROD-GW orbit formation with 1° inclination.

Time delay interferometry has been investigated for LISA much more thoroughly since 1999.[29,30] The first-generation and second-generation TDIs are proposed. In the first generation TDIs, static situations are considered. While in the second generation TDIs, motions are compensated to a certain degree. The two configurations considered above (Fig. 4) belong to the first generation TDI configurations. We shall not review more about these historical developments here, but the readers may refer to the excellent review by Tinto and Dhurandhar[30] for comprehensive treatment.

### 4.2. First generation time delay interferometry

The 1st-generation TDIs include Sagnac (α, β, γ), Unequal-arm Michelson (X, Y, Z), Relay (U, V, W), Beacon (P, Q, R), and Monitor (E, F, G) configurations. The geometric representation of these TDIs according to Vallisneri[49] is shown in Fig. 5. Each type has three sub-types based on the different initial points of spacecraft. The path length differences for the Michelson-X TDI configuration and Sagnac-α TDI configuration are shown in Fig. 4 of Section 4.1 for ASTROD-GW orbit formation with 1° inclination. Those for Relay-U, Beacon-P and Monitor-E are shown in Fig. 6.

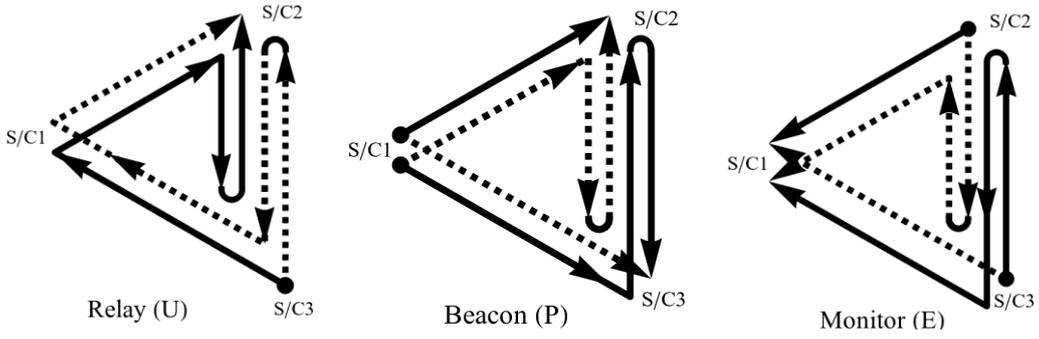

**Fig. 5.** Interference paths of the 1st-generation time-delay interferometry for Relay (U), Beacon (P) and Monitor E.

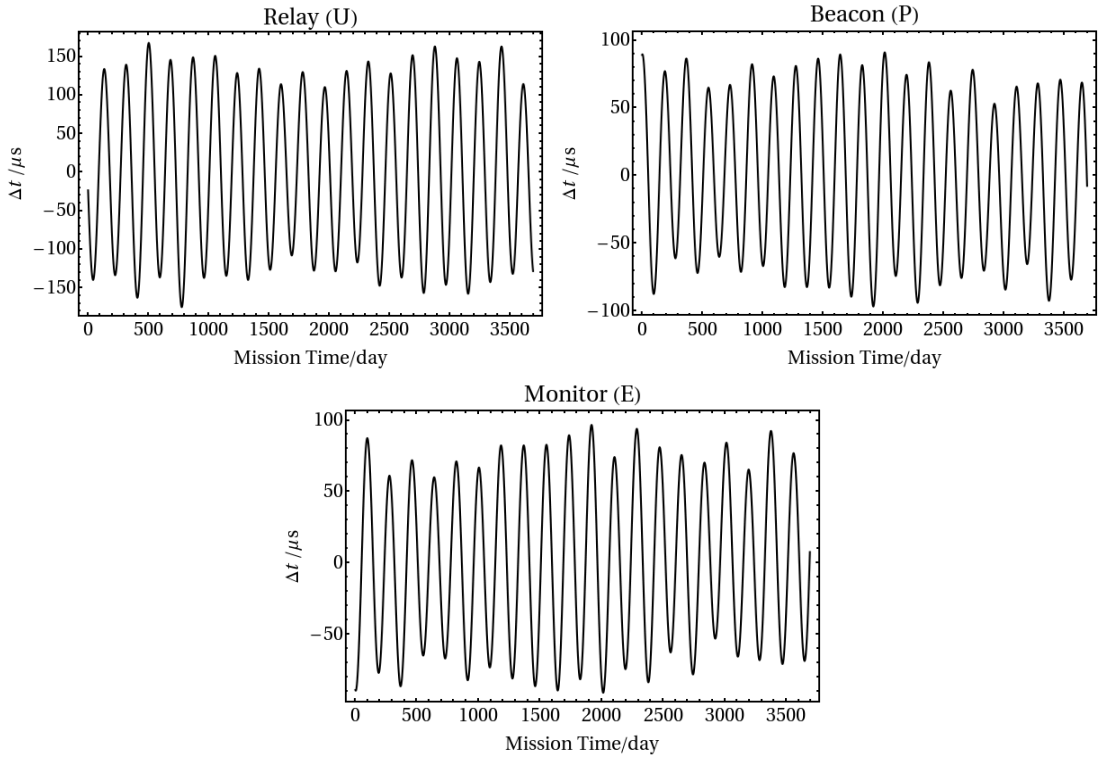

**Fig. 6.** Path length differences between two optical paths of the Relay-U, Beacon-P and Monitor-E TDI configurations for ASTROD-GW orbit formation with 1° inclination.

**Table 4.** Compilation of the rms path length differences of various first generation TDI configurations for various degrees of ASTROD-GW formation inclination (0°, 0.5°, 1°, 1.5°, 2°, 2.5° and 3°) with respect to the ecliptic plane.

| TDI configuration | ASTROD-GW TDI path difference $\Delta L$ | | | | | | |
|---|---|---|---|---|---|---|---|
| | 0°[μs] | 0.5°[μs] | 1°[μs] | 1.5°[μs] | 2.0 [μs] | 2.5 [μs] | 3.0 [μs] |
| Michelson-X | 24 | 34 | 111 | 249 | 443 | 692 | 997 |
| Michelson-Y | 18 | 35 | 113 | 251 | 444 | 692 | 996 |
| Michelson-Z | 23 | 36 | 115 | 253 | 447 | 697 | 1002 |
| Sagnac-α | 257610 | 257590 | 257531 | 257432 | 257293 | 257115 | 256898 |
| Sagnac-β | 257608 | 257588 | 257529 | 257431 | 257294 | 257118 | 256902 |
| Sagnac-γ | 257607 | 257588 | 257530 | 257432 | 257297 | 257122 | 256909 |
| Relay-U | 17 | 31 | 100 | 219 | 387 | 602 | 866 |
| Relay-V | 18 | 29 | 96 | 216 | 383 | 598 | 861 |
| Relay-W | 21 | 30 | 98 | 217 | 385 | 602 | 867 |
| Beacon-P | 12 | 17 | 56 | 125 | 222 | 346 | 499 |
| Beacon-Q | 9 | 18 | 57 | 136 | 222 | 346 | 498 |
| Beacon-R | 12 | 18 | 58 | 127 | 224 | 349 | 501 |
| Monitor-E | 12 | 17 | 56 | 125 | 222 | 346 | 499 |
| Monitor-F | 9 | 18 | 57 | 136 | 222 | 346 | 498 |
| Monitor-G | 12 | 18 | 58 | 127 | 224 | 349 | 501 |

## 4.3. Second generation TDIs in the case of one interferometric detector with two arms

As in our previous paper,[34] we calculated the TDI path length differences for the second-generation TDIs in the case of one detector with two arms obtained by Dhurandhar et al.[30] These configurations are listed in degree-lexicographic order as follows:

(I) $n = 1$, $[ab, ba] = abba - baab$,

(II) $n = 2$, $[a^2b^2, b^2a^2]$, $[abab, baba]$; $[ab^2a, ba^2b]$,

(III) $n = 3$, $[a^3b^3, b^3a^3]$, $[a^2bab^2, b2aba^2]$, $[a^2b^2ab, b^2a^2ba]$, $[a^2b^3a, {}^2a^3b]$,

$[aba^2b^2, bab^2a^2]$, $[ababab, bababa]$, $[abab^2a, baba^2b]$, $[ab^2a^2b, ba^2b^2a]$,

$[ab^2aba, ba^2bab]$, $[ab^3a^2, ba^3b^2]$, lexicographic (binary) order.

We take a common receiving time epoch for both beams and calculate the path differences. Figures 7 shows the numerical results for the $n = 1$, and $n = 2$ TDI configurations for the ASTROD-GW orbit configuration with 1° inclination. Table 4 compiles the rms path length differences of all TDI configurations listed above for various degrees of ASTROD-GW formation inclination (0°, 0.5°, 1°, 1.5°, 2°, 2.5° and 3°) with respect to the ecliptic plane.

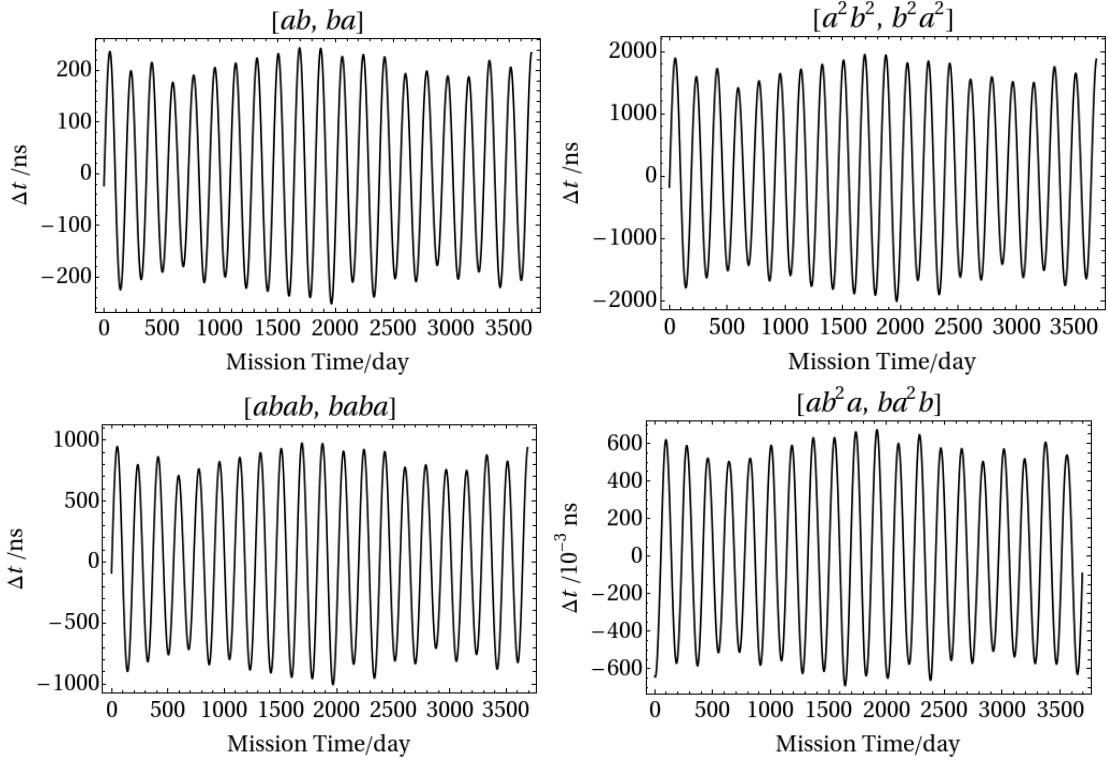

**Fig. 7.** Path length differences between two optical paths of second generation TDIs n =1 and n =2 TDI configurations for ASTROD-GW orbit formation with 1° inclination.

**Table 5.** Compilation of the rms path length differences of various TDI configurations in the case of one interferometric detector with two arms (vertex at S/C1) for various degrees of ASTROD-GW formation inclination (0°, 0.5°, 1°, 1.5°, 2°, 2.5° and 3°) with respect to the ecliptic plane. [Nominal ASTROD-GW arm length: 260 Gm]

| TDI configuration | | ASTROD-GW TDI path difference $\Delta L$ | | | | | | |
|---|---|---|---|---|---|---|---|---|
| | | 0 [ns] | 0.5 [ns] | 1 [ns] | 1.5 [ns] | 2.0 [ns] | 2.5 [ns] | 3.0 [ns] |
| $n=1$ | $[ab, ba]$ | 22 | 41 | 152 | 342 | 608 | 951 | 1370 |
| $n=2$ | $[a^2b^2, b^2a^2]$ | 173 | 328 | 1209 | 2729 | 4862 | 7605 | 10957 |
| | $[abab, baba]$ | 87 | 164 | 605 | 1365 | 2431 | 3803 | 5479 |
| | $[ab^2a, ba^2b]$ | 0.0536 | 0.112 | 0.417 | 0.942 | 1.68 | 2.63 | 3.78 |
| $n=3$ | $[a^3b^3, b^3a^3]$ | 582 | 1106 | 4079 | 9208 | 16407 | 25667 | 36980 |
| | $[a^2bab^2, b^2aba^2]$ | 453 | 860 | 3172 | 7162 | 12761 | 19964 | 28762 |
| | $[a^2b^2ab, b^2a^2ba]$ | 323 | 615 | 2266 | 5116 | 9115 | 14260 | 20544 |
| | $[a^2b^3a, ba^2a^3b]$ | 194 | 369 | 1360 | 3070 | 5469 | 8556 | 12327 |
| | $[aba^2b^2, bab^2a^2]$ | 323 | 615 | 2266 | 5116 | 9115 | 14260 | 20545 |
| | $[ababab, bababa]$ | 194 | 369 | 1360 | 3070 | 5469 | 8556 | 12327 |
| | $[abab^2a, baba^2b]$ | 65 | 123 | 454 | 1024 | 1823 | 2852 | 4109 |
| | $[ab^2a^2b, ba^2b^2a]$ | 65 | 123 | 454 | 1024 | 1823 | 2852 | 4109 |

| | | | | | | | |
|---|---|---|---|---|---|---|---|
| $[ab^2aba, ba^2bab]$ | 65 | 123 | 454 | 1024 | 1823 | 2852 | 4109 |
| $[ab^3a^2, ba^3b^2]$ | 194 | 369 | 1360 | 3070 | 5469 | 8556 | 12327 |

## 5. Discussions

We have optimized a set of 10-year inclined ASTROD-GW science mission orbits numerically using ephemeris framework starting at June 21, 2035, including cases of inclination angle is 0° (no inclination), 0.5°, 1.0°, 1.5°, 2.0°, 2.5° and 3.0°. The purpose of inclined orbit is to resolve the binary reflection ambiguity and to enhance the polar resolution. We calculate optical path length difference of various first-generation and second-generation TDIs. The original ASTROD-GW path length difference requirement is 1.5 μs; this is equivalent to rms requirement of 1 μs. (From our calculation the absolute value of maximum difference is 1.4 ~ 1.5 times the rms in 10 years except in the Sagnac TDI configuration. If Sagnac configuration needs to be used to extract GW information, more requirements are needed. We will address to this issue in the future.) Compared to this original ASTROD-GW path length difference requirement of 1 μs rms, we have the following:

(i) All the second-generation TDIs considered for one-detector case with no inclination and with 0.5° inclination satisfy this requirement (except in the marginal case $[a^3b^3, b^3a^3]$, the rms path length is 1.106 μs). (Table 5)

(ii) For the case of 1.0° inclination, of the 14 second-generation TDIs calculated, 6 satisfy the requirement. (Table 5)

(iii) For the cases of 1.5° to 3°, the requirement needs to be relaxed by 10 to 37 times; that is the laser frequency stabilization noise need to be suppressed by this additional factor. (Table 5)

(iv) Among the first-generation TDIs considered, the requirement for unequal arm Michelson, Relay, Beacon and Monitor needs to be relaxed by 2-3 orders. (Table

4)

(v) Experimental demonstration of TDI in laboratory for LISA has been implemented in 2010.[51] eLISA and the original ASTROD-GW TDI requirement are based on LISA requirement, and hence also demonstrated. With the present pace of development in laser technology, the laser frequency noise requirement is expected to be able to compensate for 2-3 order TDI requirement relaxation in 20 years.

(vi) X-configuration TDI sensitivity for GW sources has been studied extensively for eLISA.[18] It satisfies the present technological requirements well. With enhanced laser technology expected, it would also be good to study for ASTROD-GW. The study for GW sensitivity and GW sources for other first-generation and second-generation TDIs would also be encouraged.